\documentclass{nature}

\usepackage{amsmath}
\usepackage{amssymb}
\usepackage{lscape}
\usepackage{commath}
\usepackage{enumitem} 
\usepackage{soul}
\setstcolor{red}
\usepackage{caption}
\usepackage{setspace}
\usepackage{graphicx}
\usepackage[labelformat=simple,subrefformat=simple]{subcaption}

\usepackage{totcount}

\usepackage{hyperref}

\usepackage{xcolor}
\usepackage{multirow}

\usepackage{pdfpages}

\usepackage[framemethod=TikZ]{mdframed}
\usepackage{wrapfig}

\definecolor{Gray}{gray}{0.75}

\newmdenv[backgroundcolor=Gray, leftmargin = 0pt, rightmargin = 0pt, linewidth = 0pt, roundcorner = 2 pt, innerleftmargin=5pt, innerrightmargin=5pt, innertopmargin=5pt, innerbottommargin=5pt]{Frame}

\title{
\begin{center}
    Surface Optimisation Governs the Local Design of Physical Networks
\end{center}
}

\begin{document}
\maketitle
\vspace{-5em}
\small\begin{center}
Xiangyi~Meng$^{1,2,3}$, Benjamin~Piazza$^{3}$, Csaba~Both$^{3}$, Baruch~Barzel$^{3,4}$, Albert-L{\'a}szl{\'o}~Barab{\'a}si$^{3,5,6\ast}$\\
\end{center}
\vspace{-3em}
\begingroup
\fontsize{10}{7}\selectfont{\begin{center}
$^{1}$Department of Physics, Applied Physics, and Astronomy, Rensselaer Polytechnic Institute, Troy, New York 12180, USA;
$^{2}$Network Science and Technology Center, Rensselaer Polytechnic Institute, Troy, NY 12180, USA;  
$^{3}$Network Science Institute and Department of Physics, Northeastern University, Boston, Massachusetts 02115, USA; $^{4}$Department of Mathematics, Bar-Ilan University, Ramat Gan, 5290002, Israel; $^{5}$Channing Division of Network Medicine, Department of Medicine, Brigham and Women’s Hospital, Harvard Medical School, Boston, Massachusetts 02115, USA; 
$^{6}$Department of Network and Data Science, Central European University, Budapest 1051, Hungary.
\end{center}}
\endgroup

\begin{abstract}

The brain’s connectome and the vascular system are examples of physical networks whose tangible nature influences their structure, layout, and ultimately their function. The material resources required to build and maintain these networks have inspired decades of research into wiring economy, offering testable predictions about their expected architecture and organisation. Here we empirically explore the local branching geometry of a wide range of physical networks, uncovering systematic violations of the long-standing predictions of length and volume minimisation. This leads to the hypothesis that predicting the true material cost of physical networks requires us to account for their full three-dimensional geometry, resulting in a largely intractable optimisation problem. We discover, however, an exact mapping of surface minimisation onto high-dimensional Feynman diagrams in string theory, predicting that with increasing link thickness, {a locally tree-like network} undergoes a transition into configurations that can no longer be explained by length minimisation. Specifically, surface minimisation predicts the emergence of trifurcations and branching angles in excellent agreement with {the local tree organisation of} physical networks across a wide range of application domains. Finally, we predict the existence of stable orthogonal sprouts, which not only are prevalent in real networks but also play a key functional role, improving synapse formation in the brain and nutrient access in plants and fungi.

\end{abstract}

\clearpage

The vascular system and the brain are examples of physical networks, that differ from the networks typically studied in network science due to the tangible nature of their nodes and links, which
{are made of}
material resources and constrain their layout. 
The importance of these {material} factors has been noted in many disciplines:  As early as 1899, Ramón~y~Cajal suggested that we must consider the laws conserving the ‘wire’ volume to explain neuronal design\cite{hist-nerv-syst}, and in 1926 Cecil D. Murray applied {volume} minimisation principles to vascular networks, deriving the branching principles known as Murray's law\cite{murray-law_m26}. 
Today, wiring optimisation is used to account for the morphology and the layout of a wide range of physical systems\cite{phys-net_dmb18,phys-net-knot_ldb21}, from the distributions of neuronal branch sizes\cite{steiner-neuron_bkfbek10} and lengths\cite{brain-netw_mervektk13} to the morphology of plants\cite{steiner-plant_zmq01}, the structure\cite{steiner-transp_d06} and flow\cite{netw-transport_bcgldd24} in transportation networks, the layout of supply networks\cite{optim_bmr99}, the wiring of the Internet\cite{power-law-optim_dsbcbk07}, or the shape of inter-nest trails built by Argentine ants\cite{steiner-ant_lrinsmb11}, {and the design of 3D printed tissues with functional vasculature\cite{vasc_srhhsdsmspswfssm25}.}

The starting point of wiring economy approaches is the optimal wiring hypothesis, which conceptualises physical networks as a set of connected one-dimensional wires whose total length is minimised\cite{wire-min_cs99,wire-min_css02,
not-steiner_kscks12}. 
The optimal wiring in this case is exactly predicted by the Steiner graph\cite{steiner-tree-problem,steiner_r77,steiner_w87,steiner_adhblp22}.  
However, the lack of high-quality data on physical networks has limited the systematic testing of the Steiner predictions to single neuron branches{\cite{branch-angle_c92}} and ant tunnels\cite{steiner-ant_lrinsmb11}, and offered at best mixed evidence of their validity{\cite{branch-angle_z76,branch-angle_c92}}.
Yet, data availability has dramatically improved in the past few years, thanks to advances in microscopy and 3D reconstruction techniques, offering access to the detailed 3D  reconstruction of physical networks ranging from high-resolution layouts of brain connectomes\cite{insect_wpbppkfakvalrbcfhpvcz23,mouse_cbbbbbbbcccccdefffgggrhhjjjjkkkkkklllmmmmrmmmmmmnoppppprrrrrsmssssssstttttwwwwwwwwwwxyyyyyyxz23,h01-human_scjbpwbslwmskdslltcbfkfwwcallwppjl24} to vascular networks\cite{vasc-model-repos_woj13}, or the structure of coral trees\cite{3d-coral}. 
Here, we take advantage of these experimental advances to explore in a quantitative manner the role of wiring optimisation in shaping the {local morphology} of physical networks. 
We begin by documenting systematic deviations from {both the Steiner predictions\cite{steiner-tree-problem} and volume optimisation\cite{murray-law_m26,branch-angle_z76,branch-angle_c92}}, failures that we show to be rooted in the hypothesis that approximates the cost of physical networks as the sum of their link lengths\cite{wire-min_cs99,wire-min_css02,
not-steiner_kscks12} {or as simple cylinders\cite{branch-angle_c92,  branch-angle_z76}}. Indeed, the links of real physical networks are inherently three-dimensional, prompting us to hypothesise that their true material cost must also consider surface constraints. 
{Building on previous analyses that introduced volumetric constraints\cite{murray-law_m26,branch-angle_z76,branch-angle_c92}, here we successfully account for the local surface morphology, ensuring that when links intersect, they morph together continuously and smoothly, free of singularities, as dictated by the physicality of their material structure.} 
To achieve this we map {the local tree structure of} physical networks into two-dimensional manifolds, arriving at a numerically intractable surface and volume minimisation problem. 
We discover, however, a formal mapping between surface minimisation and high-dimensional Feynman diagrams, that allows us to take advantage of a well-developed string-theoretical toolset\cite{witten-cubic_w86,string-theor-quadr-differ_c88,string-theor-quadr-differ_sz89} to predict the basic characteristics of minimal surfaces.  
We find that surface minimisation can not only account for the empirically observed discrepancies from the Steiner predictions, but offers testable predictions on the degree distribution and the angle asymmetry of physical networks, that we can falsify, offering a crucial window into the design principles of physical networks. 

\section*{Results}

\begin{figure}[p!]
	\centering

    \centering
    \hspace{20pt}
    \begin{minipage}[b]{121.5pt}	
    \begin{minipage}[b]{121.5pt}	
		{\subcaption{\label{fig_principle_no}}\includegraphics[width=121.5pt]{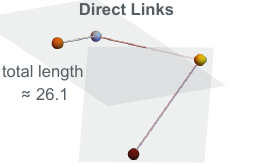}}
	\end{minipage}
    
    \begin{minipage}[b]{121.5pt}	
		{\subcaption{\label{fig_principle}}\includegraphics[width=121.5pt]{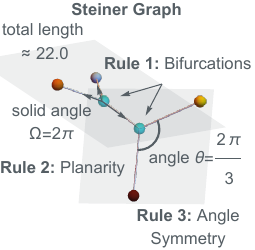}}
	\end{minipage}
    \end{minipage}
    \hspace{30pt}
    \begin{minipage}[b]{20pt}	
		{\subcaption{\label{fig_neuron_demo}}\vspace{72mm}}
    \hspace{20pt}
	\end{minipage}
    \begin{minipage}[b]{121.5pt}	
		{\includegraphics[width=121.5pt]{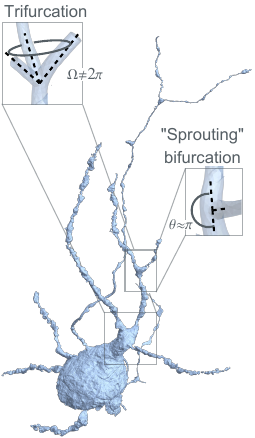}}
	\end{minipage}

    \vspace{2mm}
    
    \begin{minipage}[t]{10pt}
    \hspace{10pt}
    \end{minipage}
    \begin{minipage}[t]{10pt}
    \hspace{10pt}
    \subcaption{\label{fig_trifurcation_table}\vspace{5pt}
    }
    \end{minipage}
    \begin{minipage}[t]{345pt}
    \centering
    \vspace{-2mm}
    \definecolor{titleColor}{rgb}{0.334,0.370,0.392}
    \color{titleColor}
    \fontencoding{T1}
    \fontsize{8}{10}\fontfamily{phv}\selectfont
    \textbf{Steiner Rule 1: Bifurcations ($k=3$)}
    \begin{tabular}{|c|ccccc|}
		\hline\hline
        \noalign{\smallskip}
        & \multicolumn{5}{p{78mm}|}{Number of Nodes with Degree $k$}\\
        \hspace{2mm} Physical Network \hspace{2mm} & $3$ (bifurcation) & $4$ (trifurcation) & $5$ & $6$ & $\cdots$\\
		\noalign{\smallskip}
        \hline
		\noalign{\smallskip}
        {Human neuron} & {215,957 (76\%)} & {47,920 (17\%)} & {13,924} & {4,681}  &$\cdots$ \\
        {Fruit fly neuron} &  4,660 (92\%) & 329 {(6.5\%)} & 61 &13 &  $\cdots$ \\
        Blood vessel & 1,570 (91\%) &144 {(8.3\%)} & 17 & 3  & $\cdots$ \\
        Tropical tree & 1,512 (77\%)
        & 348 {(18\%)} & 83 & 20 & $\cdots$ \\
        Coral & 1,803 (89\%) &177 {(8.7\%)}  & 42 & 9 & $\cdots$ \\
        Arabidopsis & 791  (94\%) & 39 {(4.6\%)} & 12 & 1 & $\cdots$ \\
		\noalign{\smallskip}
        \hline\hline
	\end{tabular}
    \end{minipage}

    \vspace{2mm}

    \begin{minipage}[b]{172.6pt}	
    {\subcaption{\label{fig_planar_bi}}\includegraphics[width=172.6pt]{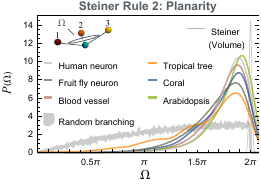}}
	\end{minipage}
    \begin{minipage}[b]{172.6pt}	
    {\subcaption{\label{fig_120}}\includegraphics[width=172.6pt]{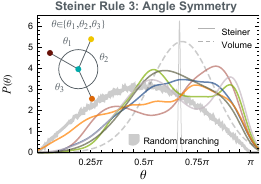}}
	\end{minipage}
    
    \caption{\label{fig_motivation}
\textbf{Real physical networks versus {length and volume optimisation} predictions}. \textbf{(\subref{fig_principle_no})}~Physical networks aim to connect spatially distributed nodes (coloured) with physical links in three dimensions. If we connect nodes directly, the wiring cost (total link length) is $\approx 26.1$.
 \textbf{(\subref{fig_principle})}~The Steiner graph minimises the wire length by permitting intermediate nodes (green), resulting in the total wire length $\approx 22.0$. The Steiner graph offers three predictions:\ 
 Rule 1.\ All branching instances are bifurcations with degree $k = 3$. 
 Rule 2.\ Bifurcations are all planar, having a solid angle of $\Omega = 2\pi$. 
 Rule 3.\ The angles between adjacent links are $\theta = 2\pi/3$.
 {Volume optimisation, which generalises links as simple cylinders of varying thickness, preserves Rules 1 and 2 and predicts a broader distribution for $\theta$, peaked around $2\pi/3$. }\textbf{(\subref{fig_neuron_demo})}~A neuron of the human connectome, demonstrating the violations of the Steiner rules. In the top inset, we highlight a trifurcation ($k = 4$) violating Rule 1. 
 We also highlight a non-symmetric branching angle,
 in which links sprout out perpendicularly (right inset), breaking Rule 3.
 \textbf{(\subref{fig_trifurcation_table})}~The percentage of $k = 4$ nodes across our six empirical {locally tree-like} physical networks. We observe $\sim 15\%$ of the nodes violating Steiner Rule 1.
 \textbf{(\subref{fig_planar_bi})}~The probability density $P(\Omega)$ vs.~$\Omega$ as obtained from all bifurcations ($k=3$) in our empirical network ensemble (coloured solid lines). {The observed density functions are more prone to Steiner Rule 2 (grey vertical line) than to random branching without optimisation (grey thick line)}.
 \textbf{(\subref{fig_120})}~The probability density $P(\theta)$ vs.~$\theta$ as obtained from all bifurcations (coloured solid lines). Once again, we observe a clear discrepancy from Steiner (grey vertical line),
 and a tendency towards random branching (grey thick line) {or volume optimisation of cylindrical links with random thickness (grey dashed line).}}
\end{figure}

The Steiner graph problem\cite{steiner-tree-problem} begins with $M$ spatially distributed nodes (Fig.~\ref{fig_principle_no}), with 
the task of connecting these nodes via the shortest possible links. 
The key insight of the Steiner solution is that by adding intermediate nodes to serve as branching points (Fig. \ref{fig_principle}), the obtained link length can be shorter than any attempt to connect the nodes directly\cite{steiner-tree-problem} (Fig. \ref{fig_principle_no}).  
{While for arbitrary $M$} the Steiner problem is  {NP-hard}, for $M=4$ we can get an exact solution,  resulting in a globally optimal Steiner graph which is characterised by three strict {local} rules (Fig.~\ref{fig_principle}): \emph{(1)\ Bifurcation only}. 
All branching instances represent bifurcations, in which a single link splits into two daughter links. 
Consequently, all intermediate nodes have degree $k = 3$, and higher degree nodes ($k > 3$) are forbidden. 
\emph{(2)\ Planarity.} At a bifurcation all three links are embedded in the same plane ($\Omega = 2\pi$); \emph{(3)\ Angle symmetry}.\ All three branches of a bifurcation form the same angle $\theta = 2\pi/3$ with each other.

To test the validity of the {local predictions of the Steiner solution}, we collected 3D resolved data of six classes of physical networks (Supplementary Section~1): 
(i) {\textit{Human neurons}}\cite{h01-human_scjbpwbslwmskdslltcbfkfwwcallwppjl24} (also in Fig.~\ref{fig_neuron_demo}),
(ii) {\textit{Fruit fly neurons}}\cite{hemibrain-fruit-fly_sxjlthhsmsbchkuzabbdkkklllnootibgwssnkabbbcccddeffffhjkkkllmmmlmnoooppppprrrrrsssssssssstttttwyhlprjcjisgmrhjp20},
(iii) \textit{Human vasculature}\cite{vasc-model-repos_woj13},
(iv) \textit{Tropical trees} from moist forests\cite{lucid-tropical-tree_gtlbharmgdmbc18},
(v) \textit{Corals} of multiple species\cite{3d-coral}, 
(vi) \textit{Arabidopsis} at different growth stages\cite{arabidopsis_phwcc21}. 
{As wiring optimisation relies on the skeleton representations of physical networks, we confirmed that our test of Steiner's prediction is not sensitive to the choice of the particular skeletonisation algorithm (Supplementary Section~1).}
To examine the validity of Rule 1 {(bifurcation only)}, we extracted the degree distribution of each skeletonised network.  
In agreement with the Steiner principle {(an outcome also predicted by volume optimisation of simple cylinders\cite{branch-angle_z76, branch-angle_c92})}, we observe a prevalence of $k = 3$ nodes, accounting for example $79\%$ of the nodes in the human {neurons} and for $94\%$ in arabidopsis. 
Yet, we also observe a significant number of trifurcations ($k = 4$), and a few even higher-degree ($k = 5,6$) nodes (Fig.~\ref{fig_trifurcation_table}), violating {both} the Steiner {and volume optimisation} prediction\cite{trifurc_p79s,trifurc_p79f}.
Note that due to errors in skeletonising a physical motif, two closely spaced bifurcations may be mistakenly identified as a trifurcation, or conversely, a trifurcation may be incorrectly perceived as two bifurcations\cite{trifurc_mthwl17}. We therefore 
verified that the observed high-degree nodes ({as demonstrated in} Fig.~\ref{fig_neuron_demo}) cannot be attributed to resolution limits (Supplementary Section~1).

To examine the validity of Rule 2 {(planarity), which predicted by both Steiner and volume optimisation}, we quantified the planarity for each bifurcation ($k=3$)
by measuring the probability  $P(\Omega)$ that the three links span a solid angle $\Omega$. 
We find that in all studied networks $P(\Omega)$ is strongly peaked at a solid angle that is smaller than $\Omega = 2\pi$, which is necessary (and sufficient) for planarity (Fig.~\ref{fig_planar_bi}). 
Finally, to test the validity of Rule 3 (angle symmetry), we extracted the pairwise angles ($\theta_1, \theta_2, \theta_3$) between the links at each bifurcation, measuring the probability density $P(\theta)$. 
As Fig.~\ref{fig_120} indicates, none of the six classes of real networks have a peak at the predicted $\theta=2\pi/3$, but instead the branching angles are broadly distributed, an asymmetry violating the Steiner prediction. 
{Note that $P(\theta)$ predicted by volume optimisation is also peaked around $\theta=2\pi/3$, but it can account for a broader range of branching angles thanks to the fact that links can have varying thickness\cite{branch-angle_z76, branch-angle_c92}}.

Taken together, while we see the signature of the Steiner theorem {and volume optimisation} in the prevalence of $k = 3$ nodes, the optimal wiring hypothesis is unable to account for the existence of $k > 3$ nodes, the prevalence of non-planar bifurcations, and the lack of $\theta = 2\pi/3$ symmetry, results that question the validity of the optimal wiring hypothesis for physical networks.

\subsection{Beyond wires---physical networks as manifold.} 
\begin{figure}[p!]
	\centering
    \begin{minipage}[b]{25pt}
    {\subcaption{\label{fig_link_geom}}\vspace{148pt}}
    \hspace{25pt}
    \end{minipage}
    \begin{minipage}[b]{110pt}	
        {\includegraphics[width=95pt]{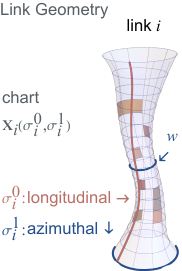}}
    \end{minipage}
    \hspace{65pt}
    \begin{minipage}[b]{15pt}
    {\subcaption{\label{fig_node_geom}}\vspace{148pt}}
    \hspace{15pt}
    \end{minipage}
    \begin{minipage}[b]{110pt}	
        {\includegraphics[width=95pt]{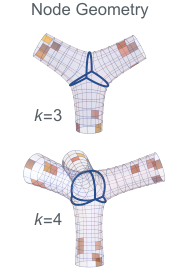}}
    \end{minipage}
    \hspace{45pt}
    \begin{minipage}[b]{25pt}
    {\subcaption{\label{fig_feynman}}\vspace{148pt}}
    \hspace{25pt}
    \end{minipage}
    \begin{minipage}[b]{110pt}	
        {\includegraphics[width=95pt]{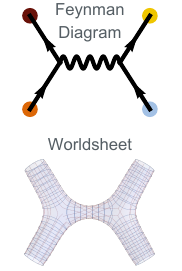}}
    \end{minipage}
    \hspace{25pt}
    
    \caption{\label{fig_phys_net}
	\textbf{Physical network manifold}.\ 
\textbf{(\subref{fig_link_geom})}~In a physical network the links are represented by charts, with a manifold morphology $\mathbf{X}_i(\boldsymbol{\sigma}_i)$. Each chart $i$ is described by its local coordinate system $\boldsymbol{\sigma}_i$. The natural parametrisation of a surface is provided by the longitudinal ($\sigma_i^0$,~red) and azimuthal ($\sigma_i^1$,~blue) coordinates.
The minimum circumference around a link is denoted by $w$, measured along a path in the azimuthal direction.
\textbf{(\subref{fig_node_geom})}~The intersections between the links define the geometry around the nodes. The local charts must be stretched and expanded to ensure a smooth and continuous patching at their boundaries (blue lines), guaranteeing that $\boldsymbol{\sigma}_i=(\sigma_i^0,\sigma_i^1)$ match perfectly with $\boldsymbol{\sigma}_j=(\sigma_j^0,\sigma_j^1)$ at the $i,j$ intersection. 
\textbf{(\subref{fig_feynman})}~A Feynman diagram (top) describes the interactions between elementary particles in field theory. In string theory, Feynman diagrams are smooth and continuous manifolds in higher dimensions (bottom) known as a worldsheet, that translate the discrete diagram on the top into the integrable object at the bottom.  An exact mapping of the surface minimisation problem [Eqs.~\eqref{eq_s_2d}~and~\eqref{eq_constraint}] to these higher dimensional worldsheets {allows} us to map abstract geometry into a structurally consistent physical network. 
}
\end{figure}

The Steiner problem relies on the hypothesis that nature aims to minimise the total length of the links, {solving an inherently global problem.} 
However, real physical networks have rich {local} geometries (Fig.~\ref{fig_neuron_demo}), characterised by varying diameters\cite{murray-law_m26} and non-cylindrical surface morphologies. 
{Over the past century, beginning with Murray’s 1926 work, researchers have combined geometry-based volume optimisation calculations\cite{murray-law_m26,branch-angle_z76,branch-angle_c92} with algorithmic approximations to identify network configurations that satisfy the inherent system-specific constraints and align with experimental data in specific domains\cite{schreiner_computer-optimization_1993,optim-vasc_jsds22,keelan_simulated_2016}.
However, these approaches cannot account either  for the smoothness of the joints that characterise real physical networks, nor for the cost associated with deviations from a simple linear or cylindrical solution.} 
Indeed, to account for the true cost of building and maintaining these networks, we must account for the full morphology of a {locally tree-like} system, which is best described as 
a \emph{manifold} $\mathcal{M}(\mathcal{G})$ assigned to the graph $\mathcal{G}$. 
Formally, a manifold is a series of \emph{charts} 
representing local coordinate systems that, when patched together, define a global 
coordinate system, or an \emph{atlas}\cite{discret-differ-geom}. 
Previous advances related graphs to \emph{discrete} manifolds through the use of simplicial complexes, assembled to form an atlas 
of connected, discrete coordinates\cite{complex-netw-manifold_br15,complex-netw-manifold_brw15,complex-netw-manifold_br16}. 
Here, however, we aim to build smooth manifolds by formally describing each chart as a continuous surface embedded in 3D, 
whose shape is described by 3D coordinates $\mathbf{X}=\left(x,y,z\right)$, where $x(\boldsymbol{\sigma})$,  $y(\boldsymbol{\sigma})$, and $z(\boldsymbol{\sigma})$ are two-variable functions of a local, two-dimensional coordinate system, $\boldsymbol{\sigma}=\left(\sigma^0, \sigma^1\right)$ (Fig.~\ref{fig_link_geom}).
This formalism replaces the total link length in the Steiner graph (Supplementary Section~2) with the total surface area {$S_{\cal M(G)}$} (Supplementary Section~3):
\begin{equation}
\label{eq_s_2d}
S_{\mathcal{M}(\mathcal{G})} = \sum_{i=1}^{L} \int \dif^{\,2}\boldsymbol{\sigma}_i \sqrt{\det \gamma_i}.
\end{equation}
Here, $\gamma_i$ is given by
$\gamma_{i,\alpha\beta} \equiv 
({\partial \mathbf{X}_i}/{\partial\sigma^{\alpha}_i}) \cdot ({\partial \mathbf{X}_i}/{\partial\sigma^{\beta}_i})$\cite{discret-differ-geom}, characterising the infinitesimal surface area elements of each link $i$. 
To ensure that the sleeves, described by $\mathbf{X}_i(\boldsymbol{\sigma}_i)$, form a smooth manifold (Supplementary Section~4) and describe a compact physical object, they must obey several strict conditions: (i) To avoid non-physical cusps when two (or more) sleeves are sewn together, the ends of the sleeves must be perfectly aligned (Fig.~\ref{fig_node_geom}), 
(ii) Surface minimisation can collapse a link, predicting that the minimum solution requires a thinning out at mid-point (Supplementary Section~5). 
However, many real physical networks must support material flux, which requires a minimum circumference $w$ everywhere, hence surface minimisation is also subject to the \emph{functional constraint} 
{\begin{equation}
    \label{eq_constraint}
    \oint_\text{circumference} \dif l_i \ge w,
\end{equation}
where the arc length is given by
$\left(\dif l_i\right)^2=\sum\nolimits_{\alpha,\beta}\gamma_{i,\alpha \beta}\dif \sigma_i^{\alpha} \dif \sigma_i^{\beta}$.
}

We, therefore, arrive at our final optimisation problem:\ given a set of terminals (pre-determined nodes), we seek the smooth and continuous surface manifold that links all terminals via finite paths, whose circumference exceeds the pre-defined threshold $w$ and minimises the cost $S_{\mathcal{M}(\mathcal{G})}$ [Eq.~\eqref{eq_s_2d}].
At first glance, this optimisation problem is intractable, as we must compare an uncountably infinite set of circumferences, known as non-contractable closed curves\cite{partial-differ-relat}, ensuring that none of them violate Eq.~\eqref{eq_constraint} while minimising Eq.~\eqref{eq_s_2d}.
Our key methodological advance is the discovery of a direct equivalence between the network manifold minimisation problem defined above and higher-dimensional Feynman diagrams (known as pants decomposition) in string theory\cite{witten-cubic_w86,string-theor-quadr-differ_c88,string-theor-quadr-differ_sz89}.
The traditional Feynman diagram is a graph $\mathcal{G}$ that views particle trajectories as links and collisions as nodes (Fig.~\ref{fig_feynman}). 
String (field) theory generalises Feynman diagrams to two-dimensional surfaces, called the `worldsheets,' which represent the paths that strings sweep through in spacetime\cite{witten-cubic_w86,string-theor-quadr-differ_c88,string-theor-quadr-differ_sz89}. 
The smoothness of this surface guarantees that the path integral does not diverge, making it renormalisable\cite{string-theor_t09}, resulting in the Nambu--Goto action\cite{string-theor_t09} that is formally identical to Eq.~\eqref{eq_s_2d}.
{The classical solution of the Nambu--Goto action, obtained in the absence of quantum fluctuations but subject to the constraint Eq.~\eqref{eq_constraint}, is exactly the manifold $\mathcal{M}(\mathcal{G})$ we seek. 
According to Strebel's theorem, in the absence of boundary conditions, this minimal surface is exactly cylindrical.
With boundary conditions added, we can simplify Eq.~\eqref{eq_constraint} to a local constraint (Supplementary Section~5), allowing us to construct {local trees} with discrete surfaces that are optimised for both smoothness and minimality.
Numerically this is performed by the \emph{min-surf-netw} package, described in Supplementary Section~6 and shared on GitHub.}

\subsection{Degree distribution.}
\begin{figure}[p]
\vspace{-10mm}
	\centering
    \hspace{10pt}
    \begin{minipage}[t]{162.6pt}
    \vspace{8pt}
    \begin{minipage}[t]{80pt}
		\centering
		{\subcaption{\label{fig_trifur_demo}\vspace{-12mm}}
        \includegraphics[width=170pt]{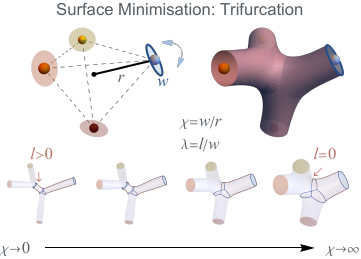}}
	\end{minipage}
    \begin{minipage}[t]{80pt}
		\centering
		\subcaption{\label{fig_trifur_demo_b}\vspace{-12mm}}
	\end{minipage}

    \begin{minipage}[t]{38.5pt}
		\centering
		\vspace{-7mm}\subcaption{\label{fig_trifur_demo_c}}
	\end{minipage}
    \begin{minipage}[t]{38.5pt}
		\centering
		\vspace{-7mm}\subcaption{\label{fig_trifur_demo_d}}
	\end{minipage}
    \begin{minipage}[t]{38.5pt}
		\centering
		\vspace{-7mm}\subcaption{\label{fig_trifur_demo_e}}
	\end{minipage}
    \begin{minipage}[t]{38.5pt}
		\centering
		\vspace{-7mm}\subcaption{\label{fig_trifur_demo_f}}
	\end{minipage}
    \end{minipage}
     \begin{minipage}[t]{172.6pt}	
		\centering
		{\subcaption{\label{fig_trifur}}\includegraphics[width=172.6pt]{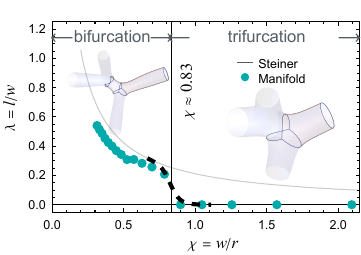}}
	\end{minipage}
	\begin{minipage}[t]{172.6pt}	
		\centering
		{\subcaption{\label{fig_trifur_logpdf}}\includegraphics[width=172.6pt]{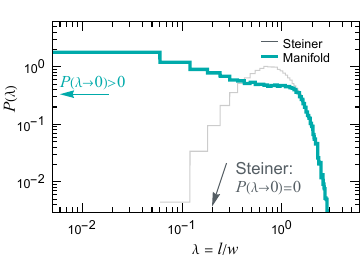}}
	\end{minipage}

   \begin{minipage}[b]{172.6pt}
		\centering
		{\subcaption{\label{fig_trifurcation_neuron_human}}\includegraphics[width=172.6pt]{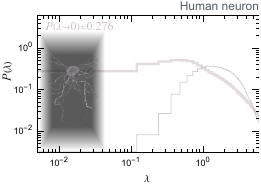}}
	\end{minipage}
    \begin{minipage}[b]{172.6pt}
		\centering
		{\subcaption{\label{fig_trifurcation_neuron_fruit_fly}}\includegraphics[width=172.6pt]{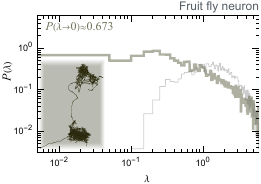}}
	\end{minipage}
    \begin{minipage}[b]{172.6pt}
		\centering
		{\subcaption{\label{fig_trifurcation_vascular}}\includegraphics[width=172.6pt]{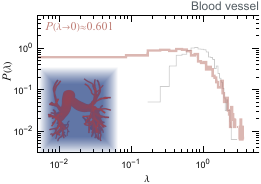}}
	\end{minipage}
    
    \begin{minipage}[b]{172.6pt}
		\centering
		{\subcaption{\label{fig_trifurcation_lucid}}\includegraphics[width=172.6pt]{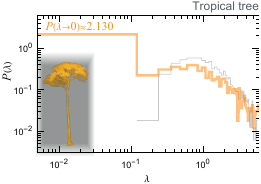}}
	\end{minipage} 
    \begin{minipage}[b]{172.6pt}
		\centering
		{\subcaption{\label{fig_trifurcation_coral}}\includegraphics[width=172.6pt]{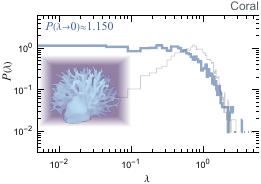}}
	\end{minipage} 
    \begin{minipage}[b]{172.6pt}
		\centering
		{\subcaption{\label{fig_trifurcation_arabidopsis}}\includegraphics[width=172.6pt]{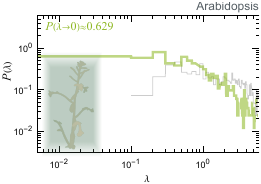}}
	\end{minipage}
 
    \caption{\label{fig_trifurcation}
	\textbf{Emergence of trifurcations.}
    \textbf{(\subref{fig_trifur_demo})}~We consider four nodes forming a perfect tetrahedral configuration with spatial length scale $r$, capturing the tetrahedron's radius. 
     \textbf{(\subref{fig_trifur_demo_b})}~We construct a physical network to link these four nodes under surface minimisation with circumference constraint $w$ (link thickness). 
    \textbf{(\subref{fig_trifur_demo_c},\subref{fig_trifur_demo_d})}~When $\chi=w/r \to 0$, the sleeves behave as one-dimensional links, and the resulting manifold is well approximated by the Steiner solution, the network featuring two $k = 3$ bifurcations. 
    \textbf{(\subref{fig_trifur_demo_e},\subref{fig_trifur_demo_f})}~As $\chi$ increases, the intermediate link $l$ becomes shorter, until, beyond a certain thickness the separation parameter $\lambda=l/w \to 0$, indicating that the two intermediate bifurcations unite into a single trifurcation with $k = 4$. 
     \textbf{(\subref{fig_trifur})}~To examine the predicted transition we plot $\lambda$ vs.~$\chi$ for the minimal surface (green). For small $\chi$ we have $\lambda> 0$, following a pattern also predicted by Steiner (grey solid line). This captures the two-bifurcation scenario predicted by length minimisation. However, at $\chi \approx 0.83$ we observe a sudden drop to $\lambda = 0$, capturing the transition from double bifurcations to a single trifurcation.   
     \textbf{(\subref{fig_trifur_logpdf})}~We examined a series of random four-node configurations within a unit cube and implicitly constructed for each a Steiner graph and a minimal surface manifold ($w = 1$). We then extracted $P(\lambda)$, capturing the probability density to observe  $\lambda$. 
     Under Steiner optimisation, $P(\lambda)$ vanishes as $\lambda \to 0$ (grey curve), capturing the fact that trifurcations are forbidden. In contrast, for surface minimisation (green curve) we have $P(\lambda \to 0) > 0$, describing a finite likelyhood to observe trifurcations. 
     \textbf{(\subref{fig_trifurcation_neuron_human})--(\subref{fig_trifurcation_arabidopsis})}~$P(\lambda)$ vs.~$\lambda$ obtained from real physical networks. In each network we collected all tetrahedral motifs in which the four external nodes are linked through two intermediate nodes, and extracted $\lambda$ between these intermediaries. Compared to Steiner's predictions (grey), the empirically observed $P(\lambda)$ (distinct colours) follows the green pattern in (\subref{fig_trifur_logpdf}), capturing a coexistence of bifurcations ($\lambda > 0$) and trifurcations ($\lambda=0)$, as predicted by surface minimisation. 
        }
\end{figure}

We start from a symmetric configuration of four terminals, laid out on the corners of a regular tetrahedron (Fig.~\ref{fig_trifur_demo}) and construct the minimal-surface network motif, {represented by a tree} that links these four nodes, with minimal link circumference $w$ (Fig.~\ref{fig_trifur_demo_b}). 
Defining the dimensionless weight parameter, $\chi = w/r$, where $r$ is the distance between the intermediate nodes, in the $\chi \to 0$ limit we have a quasi-one-dimensional configuration with long and thin links. 
In this case, the surface minimisation predictions converge to the Steiner Rules 1--3 (Fig.~\ref{fig_principle}), linking the four terminal nodes via two intermediate bifurcations with degree $k = 3$ (Fig.~\ref{fig_trifur_demo_c}~and~\ref{fig_trifur_demo_d}). 
Yet, the optimal solution also predicts that for higher $\chi$ (thicker links) the two $k=3$ nodes gradually approach each other, and that at $\chi \sim 1$ they merge into a single $k = 4$ node, resulting in trifurcation (Fig.~\ref{fig_trifur_demo_e}~and~\ref{fig_trifur_demo_f}). 
In other words, surface minimisation\cite{string-theor-quadr-differ_sz89} predicts a transition from a Steiner bifurcation to a stable trifurcation at $\chi \sim 1$, {an outcome that eluded volume optimisation as well\cite{branch-angle_z76,branch-angle_c92}.} 

To quantify this transition, we use the dimensionless separation $\lambda = l / w$ as order parameter, where $l$ is the length of the link between the two $k=3$ nodes, and using \emph{min-surf-netw} (Supplementary Section~6) we numerically generate the connecting minimal surface, allowing us to measure $\lambda (\chi)$ as a function of $\chi$.  
For small $\chi$, we have $\lambda > 0$, predicting that the two $k = 3$ nodes are separated, in line with the Steiner prediction (Fig.~\ref{fig_trifur}). 
Yet, at $\chi \approx 0.83$ we observe a sudden drop to $\lambda = 0$, when the one-dimensional Steiner approximation breaks down, and instead surface minimisation predicts the emergence of a trifurcation ($k = 4$). 
This transition represents our first key prediction, indicating that the empirically observed $k = 4$ nodes in {locally tree-like} physical networks represent a stable configuration predicted by {local} surface optimisation.

To generalise our approach, we place the four terminals randomly in a unit cube, and run multiple configurations to extract the probability density $P(\lambda)$.
For $\chi = 0$ (corresponding to $w = 0$ which reduces to the Steiner problem), we find that $P(\lambda) \to 0$ for small separation $\lambda$ {(Fig.~\ref{fig_trifur_logpdf}, grey line)}, confirming the absence of trifurcations. 
In contrast, for large $\chi$ (e.g.~$w = 1$), we find that $P(\lambda \to 0)$ does not vanish {(Fig.~\ref{fig_trifur_logpdf}, green line)}, but for $\lambda = 0$ we predict a finite probability for trifurcations (Supplementary Section~7). 
Figure \ref{fig_trifur_logpdf} indicates that the density function $P(\lambda)$ offers an empirically falsifiable fingerprint of surface minimisation. 
We therefore divided each physical network into local groups of four connected links and extracted $P(\lambda)$. 
We find that each {locally tree-like} network exhibits a non-vanishing $P(\lambda\to 0)$ (Fig.~\ref{fig_trifurcation_neuron_human}--\ref{fig_trifurcation_arabidopsis}, coloured lines), representing a clear deviation from the Steiner prediction (green line) and offering direct evidence that in real networks the cost function is not linear in the link length, but it is better described by surface minimisation.

\subsection{Angle asymmetry.}
\begin{figure}[p]
	\centering
    \hspace{10pt}
    \begin{minipage}[t]{162.6pt}
    \vspace{8pt}
    \begin{minipage}[t]{80.0pt}
		\centering
		{\subcaption{\label{fig_bimodal_demo}\vspace{-12mm}}
        \includegraphics[width=170pt]{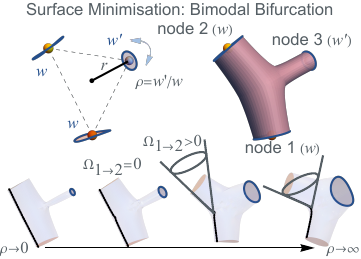}}
	\end{minipage}
    \begin{minipage}[t]{80.0pt}
		\centering
		\subcaption{\label{fig_bimodal_demo_b}\vspace{-12mm}}
	\end{minipage}

    \begin{minipage}[t]{38.5pt}
		\centering
		\vspace{-2mm}\subcaption{\label{fig_bimodal_demo_c}}
	\end{minipage}
    \begin{minipage}[t]{38.5pt}
		\centering
		\vspace{-2mm}\subcaption{\label{fig_bimodal_demo_d}}
	\end{minipage}
    \begin{minipage}[t]{38.5pt}
		\centering
		\vspace{-2mm}\subcaption{\label{fig_bimodal_demo_e}}
	\end{minipage}
    \begin{minipage}[t]{38.5pt}
		\centering
		\vspace{-2mm}\subcaption{\label{fig_bimodal_demo_f}}
	\end{minipage}
    \end{minipage}
    \begin{minipage}[t]{172.6pt}
		\centering
		{\subcaption{\label{fig_bimodal}}\includegraphics[width=172.6pt]{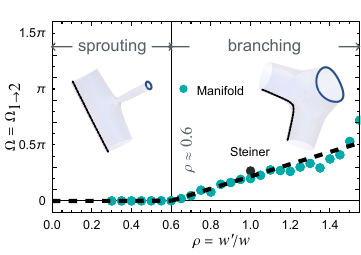}}
	\end{minipage}
    \begin{minipage}[t]{172.6pt}
        \centering
		{\subcaption{\label{fig_bimodal_sprout}}\includegraphics[width=172.6pt]{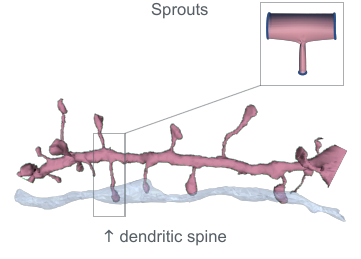}}
    \end{minipage}

    \begin{minipage}[b]{172.6pt}
		\centering
		{\subcaption{\label{fig_bimodal_bifurcation_accumulate_neuron_human_loglog}}\includegraphics[width=172.6pt]{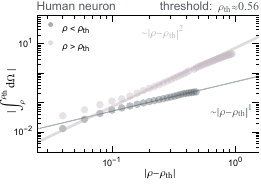}}
	\end{minipage}
    \begin{minipage}[b]{172.6pt}
		\centering
		{\subcaption{\label{fig_bimodal_bifurcation_accumulate_neuron_fruit_fly_loglog}}\includegraphics[width=172.6pt]{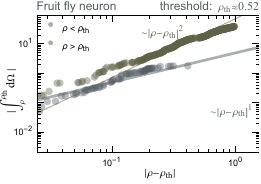}}
	\end{minipage}
    \begin{minipage}[b]{172.6pt}
		\centering
		{\subcaption{\label{fig_bimodal_bifurcation_accumulate_vascular_loglog}}\includegraphics[width=172.6pt]{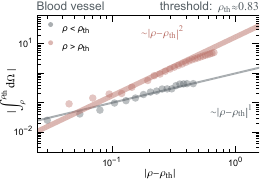}}
	\end{minipage}

    \begin{minipage}[b]{172.6pt}
		\centering
		{\subcaption{\label{fig_bimodal_bifurcation_accumulate_lucid_loglog}}\includegraphics[width=172.6pt]{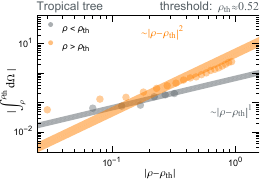}}
	\end{minipage}
    \begin{minipage}[b]{172.6pt}
		\centering
		{\subcaption{\label{fig_bimodal_bifurcation_accumulate_coral_loglog}}\includegraphics[width=172.6pt]{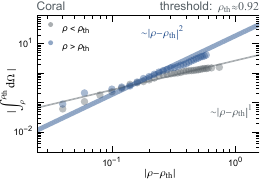}}
	\end{minipage}
    \begin{minipage}[b]{172.6pt}
		\centering
		{\subcaption{\label{fig_bimodal_bifurcation_accumulate_arabidopsis_loglog}}\includegraphics[width=172.6pt]{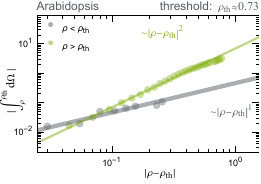}}
	\end{minipage}

    \caption{\label{fig_bimodal_bifurcation}
	\textbf{Branching versus sprouting bifurcations}. 
 \textbf{(\subref{fig_bimodal_demo})}~We start from a triangular node configuration, with $w_1 = w_2=w$ and $w_3=w^\prime$.
 \textbf{(\subref{fig_bimodal_demo_b})}~We construct the minimal surface manifold connecting the three nodes. \textbf{(\subref{fig_bimodal_demo_c},\subref{fig_bimodal_demo_d})}~For small $\rho = w^\prime / w$, the link of node $3$ is thin, and the optimal manifold favours a sprouting structure: nodes $1$ and $2$ linked through a straight line and node $3$ via a perpendicular link.
\textbf{(\subref{fig_bimodal_demo_e},\subref{fig_bimodal_demo_f})}~For large $\rho$, we find a linear relation between $\rho$ and the three-dimensional steering angle, $\Omega_{1\to2}$, related to the branching angle $\theta$ (Fig.~\ref{fig_120}) via $\Omega_{1\to2}=4 \pi \sin ^2\left[\left(\pi-\theta\right)/4\right]$. As $\rho$ increases, the bifurcation point approaches the triangle centre, and the bifurcation gradually resembles a symmetric branching.
 \textbf{(\subref{fig_bimodal})}~$\Omega_{1\to2}$ vs.~$\rho$. We observe a transition from sprouting ($\Omega = 0$) to branching ($\Omega > 0$) at $\rho \approx 0.6$. The symmetric branching observed by Steiner appears near $\rho = 1$.
 \textbf{(\subref{fig_bimodal_sprout})}~In the human connectome $92\%$, of the observed sprouts end on synapses, suggesting that neuronal systems utilise surface minimisation to form direct synaptic connections to adjacent neurons with minimal material cost. 
 \textbf{(\subref{fig_bimodal_bifurcation_accumulate_neuron_human_loglog})--(\subref{fig_bimodal_bifurcation_accumulate_arabidopsis_loglog})}~According to \textbf{(\subref{fig_bimodal})}, cumulative  
 {$|\int\nolimits_{\rho}^{\rho_\text{th}}\Omega(\rho)\dif\rho|$} should follow $\sim \left(\rho_\text{th}-\rho\right)^{1}$ for $\rho<\rho_\text{th}$  and 
 $\sim \left(\rho-\rho_\text{th}\right)^{2}$ for $\rho>\rho_\text{th}$, predictions closely followed by real physical networks.
 {Band thickness represents one standard error of the fitting.}}
\end{figure}

To understand the origin of the observed angle diversity, a violation of Rule 3 (Fig.~\ref{fig_120}), 
we assume that each link $i$ is characterised by its unique circumference constraint $w_i$. Without a loss of generality, we set $w_1 = w_2 = w$ and $w_3 = w^\prime$, and vary the ratio $\rho = w'/w$, to obtain the minimal manifold that connects nodes $1$, $2$, and $3$ (Fig.~\ref{fig_bimodal_demo}~and~\ref{fig_bimodal_demo_b}).  While Steiner's solution posits a constant steering angle $\Omega_{1 \to 2}\approx 0.3 \pi$, surface minimisation predicts two distinct regimes {separated by a threshold value $\rho_\text{th}$ (Supplementary Section~7)}: 
(1) For {$\rho>\rho_\text{th}$}, we predict the steering angle $\Omega_{1 \to 2}\approx k (\rho-\rho_\text{th})$ 
(Fig.~\ref{fig_bimodal_demo_e}~and~\ref{fig_bimodal_demo_f}), i.e.~a \emph{linear} dependence on $\rho$ (Fig.~\ref{fig_bimodal}).  This regime can therefore account for the wide range of angles observed in Fig.~\ref{fig_120}.  
(2) For {$\rho<\rho_\text{th}$}, surface minimisation makes an unexpected prediction: if links $1$ and $2$ have comparable diameters, they are expected to form a straight path (i.e.~continue with solid angle of $\Omega_{1\to2} = 0$), while the thinner link $3$ is predicted to emerge perpendicularly at $\Omega_{1\to3} \approx \Omega_{2\to3}$, consistent with an orthogonal \emph{sprouting} behaviour (Fig.~\ref{fig_bimodal_demo_c}~and~\ref{fig_bimodal_demo_d}). Note that a geometric approach predicted as early as 1976\cite{branch-angle_z76, branch-angle_c92} that the branch angles converge to $90$ degrees in the $\rho \to 0$ limit  (Supplementary Section~7). 
In contrast, our framework predicts that the $90$-degree solution is optimal for any $\rho< \rho_\text{th}$ (Fig.~\ref{fig_bimodal}). Hence, orthogonal sprouts are not singular solutions that emerge only in the $\rho \to 0$ limit\cite{branch-angle_z76,branch-angle_c92}. 
Rather, they are stable solutions of surface minimisation that remain minimal for a wide range of parameter values, and hence they should be not only observable, but prevalent in real physical networks. 

To test these predictions, we identified all bifurcation motifs in each network in our database, and then searched for branches that satisfy $w_1 = w_2 = w$. We then measured  $\Omega(\rho)=\Omega_{1 \to 2}$ in function of the empirically observed $\rho$, finding that almost all bifurcations for $\rho<\rho_\text{th}$ are sprout-like, characterised by small $\Omega(\rho)$ (Supplementary Section~7).
In Fig.~\ref{fig_bimodal_bifurcation_accumulate_neuron_human_loglog}--\ref{fig_bimodal_bifurcation_accumulate_arabidopsis_loglog}, we show the cumulative value of the observed angles in the two regimes, offering evidence that the cumulative {$|\int\nolimits_{\rho}^{\rho_\text{th}}\Omega(\rho)\dif\rho|$} follows 
$\sim \left(\rho_\text{th}-\rho\right)^{1}$ for $\rho<\rho_\text{th}$  and a quadratic behaviour 
$\sim \left(\rho-\rho_\text{th}\right)^{2}$ for  $\rho>\rho_\text{th}$, in line with the predictions of Fig.~\ref{fig_bimodal}.

The key outcome of surface minimisation is the predicted prevalence of the orthogonal sprouts, expected to emerge each time $\rho<\rho_\text{th}$. To falsify this prediction, we ask: are such sprouts really present in physical networks? 
{Note that the excess of sprouts over the expectations of length or volume optimisation was already noted in arterial systems as early as 1976\cite{branch-angle_z76}. This abundance remained unanswered, and it also remains unclear whether sprouts represent a generic feature across all physical networks, or are unique to blood vessels.}  
To address this, we first identified all bifurcations with $w_1\approx w_2$ in blood vessels, confirming that in  $25.6\%$ of the cases the third branch, independent of $\rho$, is perpendicular to the main branches, representing {an abundant sprouting behaviour. 
Yet, we find that sprouts} are not limited to the circulatory system, but are present in all studied networks,  representing $12.9\%$ of the $w_1\approx w_2$ cases in the tropical trees, $52.8\%$ in corals, $11.2\%$ in arabidopsis, $13.8\%$ in the {fruit fly neurons}, and  $18.4\%$ in the {human neurons}.  
Most importantly, some systems have learned to turn sprout behaviour to their advantage, assigning it a functional role. 
Indeed, in the human connectome we identified $4,003$ sprouts, finding that $3,911$ of these  ($98\%$) end with a synapse (Fig.~\ref{fig_bimodal_sprout}).  In other words, neuronal systems have adapted to rely on surface minimisation by using orthogonal sprouts as dendritic spines that allow them to form synapses with nearby neurons with minimal material cost. 
Similarly, roots in plants\cite{lynch2013steep} and hyphae branches in fungi\cite{harris2008branching} are known to sprout perpendicularly, allowing plants and fungi to explore a larger volume of soil for water and nutrients with minimal material expenditure.

\begin{figure}[p]
	\centering
    \begin{minipage}[b]{172.6pt}
		\centering
		{\subcaption{\label{fig_bimodal_distribution_neuron_human}}\includegraphics[width=172.6pt]{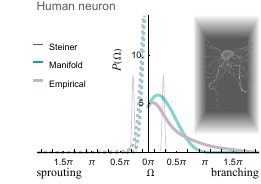}}
	\end{minipage}
    \begin{minipage}[b]{172.6pt}
		\centering
		{\subcaption{\label{fig_bimodal_distribution_neuron_fruit_fly}}\includegraphics[width=172.6pt]{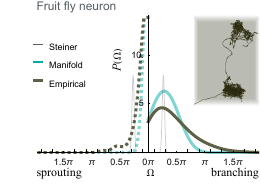}}
	\end{minipage}
    \begin{minipage}[b]{172.6pt}
		\centering
		{\subcaption{\label{fig_bimodal_distribution_vascular}}\includegraphics[width=172.6pt]{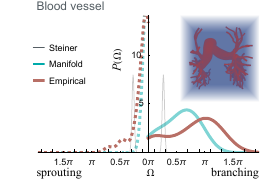}}
	\end{minipage}

    \begin{minipage}[b]{172.6pt}
		\centering
		{\subcaption{\label{fig_bimodal_distribution_lucid}}\includegraphics[width=172.6pt]{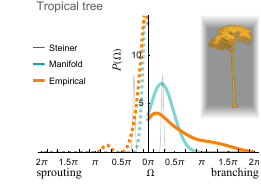}}
	\end{minipage}
    \begin{minipage}[b]{172.6pt}
		\centering
		{\subcaption{\label{fig_bimodal_distribution_coral}}\includegraphics[width=172.6pt]{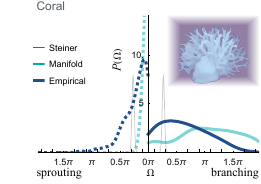}}
	\end{minipage}
    \begin{minipage}[b]{172.6pt}
		\centering
		{\subcaption{\label{fig_bimodal_distribution_arabidopsis}}\includegraphics[width=172.6pt]{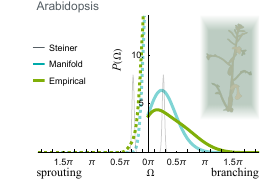}}
	\end{minipage}

    \caption{\label{fig_bimodal_distribution}
	{\textbf{Sprouting in physical networks}. We predicted and measured the branching angle distribution across six physical networks. \textbf{(\subref{fig_bimodal_distribution_neuron_human})--(\subref{fig_bimodal_distribution_arabidopsis})}~The relation of $\Omega_{1\to2}$ vs.~$\rho$ in Fig.~\ref{fig_bimodal_bifurcation} predicts distinct distributions $P(\Omega)$ based on the observed $\rho$ values in the sprouting (dashed) and branching (solid) regimes. Both distributions align with our predictions (green), violating the Steiner predictions (grey).}}
\end{figure}

The predicted relation between $\Omega(\rho)$ and $\rho$ in Fig.~\ref{fig_bimodal} leads to further falsifiable predictions for the $P(\Omega)$ angle distributions, conditioned on the empirically observed $\rho$ values. 
In the sprouting regime ($\rho<\rho_\text{th}$), we predict {$\Omega=0$}, independent of $\rho$, hence we anticipate a sharp peak of $P(\Omega)$ at $\Omega = 0$, in agreement with the empirical data (left side, sprouting regime in Fig.~\ref{fig_bimodal_distribution_neuron_human}--\ref{fig_bimodal_distribution_arabidopsis}). 
In the branching regime ($\rho>\rho_\text{th}$),  however, $P(\Omega)$ is predicted to exhibit a broad distribution with high variance, rooted in the linear behaviour of Fig.~\ref{fig_bimodal}. The empirical data support this prediction as well (right side, branching regime in Fig.~\ref{fig_bimodal_distribution_neuron_human}--\ref{fig_bimodal_distribution_arabidopsis}). In comparison, the Steiner prediction posits a sharp peak of $P(\Omega)$ independent of $\rho$ (grey thin lines in both sprouting and branching regimes in Fig.~\ref{fig_bimodal_distribution_neuron_human}--\ref{fig_bimodal_distribution_arabidopsis}).

\subsection{Discussion.}
{The 3D layout of physical networks is subject to multiple, often evolutionary-induced constraints.
For example, brain wiring is governed by developmental programmes\cite{gene-connect_bb20}, and locally guided by a complex inventory of chemo-attractants and repellents that govern an individual neuron's journey across the brain. Similarly, the vascular system must transport nutrients to all cells and is subject to multiple optimisation goals, from flow efficiency to material cost\cite{optim_wbe97}. 
Given the diversity of the processes that govern the development of physical networks, one would expect that minimisation principles are ultimately overwritten by global and functional needs\cite{thompson_growth_1992,west_scale_2017}}. In contrast, here we find that physical networks observed in a wide range of systems follow common quantifiable morphological {branching} characteristics that are well predicted by a {local} surface minimisation process. 
The robustness of our results across multiple systems indicates that  cost minimisation is a stereotypical principle that is not overwritten {by functional or global} need; rather,  development and selection likely rely upon these {local} minimisation processes to add function to a network. 
{As local optimisation does not necessarily dictate the global optimum\cite{branch-angle_c92}, functional demands may exert greater influence at larger scales\cite{optim-vasc_jsds22,vasc_srhhsdsmspswfssm25}.
For example, we find that wiring optimisation fails to correctly predict the total length of physical networks, which are on average $25\%$ longer than Steiner's prediction across all six datasets (Supplementary Section 8).}

{Additional empirical studies are needed to validate surface minimisation predictions across more complex
network structures\cite{netw-struct-dyn_blmch06}. 
Indeed, while here we focused on the universal branching characteristics {of locally tree-like structures}, construction of larger-scale structures
could reveal whether specific network types exhibit unique geometrical adaptations, such as varying link thickness and curvature, due to the networks' unique functional pressures, like flow conservation in vascular systems\cite{murray-law_m26} or  neuron placement constraints\cite{gene-connect_bb20}. 
These features are beyond the scope of our current surface minimisation framework, which predicts straight, uniform cylinders far from the branching points. 
Furthermore, loops---which we find to be absent in our datasets (Supplementary Section~8) but ubiquitous in engineered networks like traffic and power grids---represent a departure from simple wiring efficiency, hence requiring an extended analytical framework.
Such advances will open avenues to integrate crowding\cite{phys-net_dmb18,phys-net-meta_psbbaklb24}, knotting\cite{phys-net-knot_ldb21,phys-net-link_gb24}, or bundling\cite{phys-net-bundle_brpaksklb24}
of physical links, exploring their influence on the global layout. 
Such extensions could offer further insights into how networks balance efficiency with functional demands\cite{stat-phys-netw_cssggc19}, and help us understand how a global and functional organisation can emerge from local processes. They may also offer insight into differences between classes of physical networks, helping us understand which features are governed by optimisation principles, and which require additional functional considerations.}

{Future work could also compare the predicted manifold geometries directly to the observed geometric features, like surface geodesics, curvatures, and other fine details, helping reveal the degree to which the surface minimisation model reproduces the observed local geometry beyond skeletons. Indeed, we find that trifurcation junctions are consistently smooth and that their shapes strongly prefer symmetric morphology, features predicted by surface minimisation (Supplementary Section 9). This validation at the level of fine-grained geometry reinforces our framework's empirical foundation and opens avenues for more detailed comparison with the predictions.}

{Physical networks in the 3D Euclidean space can be described as either two-dimensional manifolds $\mathcal{M}(\mathcal{G})$ subject to surface minimisation, or three-dimensional objects subject to volume optimisation.  While in vascular networks the material investment is limited to the surface area of the blood vessels, for neurons, corals, and trees, an accurate accounting of the material cost must also consider the volume of the branches. The existing literature on volume optimisation assumes cylindrical links\cite{branch-angle_z76,branch-angle_c92} and fails to account for non-trivial topologies emerging at the intersections. As the \emph{min-surf-netw} algorithm exploits the string-theoretic solution, it is limited to surface minimisation.}
{Yet, the two problems are not independent: our numerical simulations indicate that for the branching processes sub-optimal surfaces also increase the volume, suggesting that the minimal surfaces correspond to close-to-optimal volumes as well (Supplementary Section~10).}
However, further work is needed to understand if a self-consistent volume optimisation could offer novel solutions and morphologies that are not predicted by our current framework, hence can further enrich our understanding of physical networks.

\newpage

\newpage
\section*{References}
\bibliography{Manifold}

\newpage
\begin{addendum}
    \item {We are grateful to Ulf H. Danielsson and Fabian Ruehle for helpful discussions regarding the string theory approach.} This research was partially supported by the NSF award No.~2243104 – COMPASS and by the European Union’s Horizon 2020 research and innovation programme No.~810115 – DYNASNET. 
    
    \item[Data Availability] 
    The dataset is available at \href{https://physical.network}{https://physical.network}
    
    \item[Code Availability] 
    The code used for this manuscript is available at \href{https://github.com/Barabasi-Lab/min-surf-netw}{https://github.com/Barabasi-Lab/min-surf-netw}
    
    \item[Author Contributions] 
    All authors contributed to the research. X.M. and A.-L.B. conceived the research. X.M., B.P., and C.B. collected and cleaned the data. X.M. and B.P. analysed the data.
    X.M. conducted theoretical analysis, designed the algorithm, and performed the simulation.
    B.B., X.M., and A.-L.B. wrote the manuscript. X.M., C.B., and A.-L.B. reviewed and edited the manuscript.

    \item[Competing Interests] A.-L.B. is the scientific founder of Scipher Medicine, Inc., which applies network medicine to biomarker development.
    
    \item[Correspondence] Correspondence and requests for materials should be addressed to A.-L.B.\newline (email: a.barabasi@northeastern.edu).

\end{addendum}

\pagenumbering{gobble}

\newpage

\newpage

\newpage

\includepdf[pages=-]{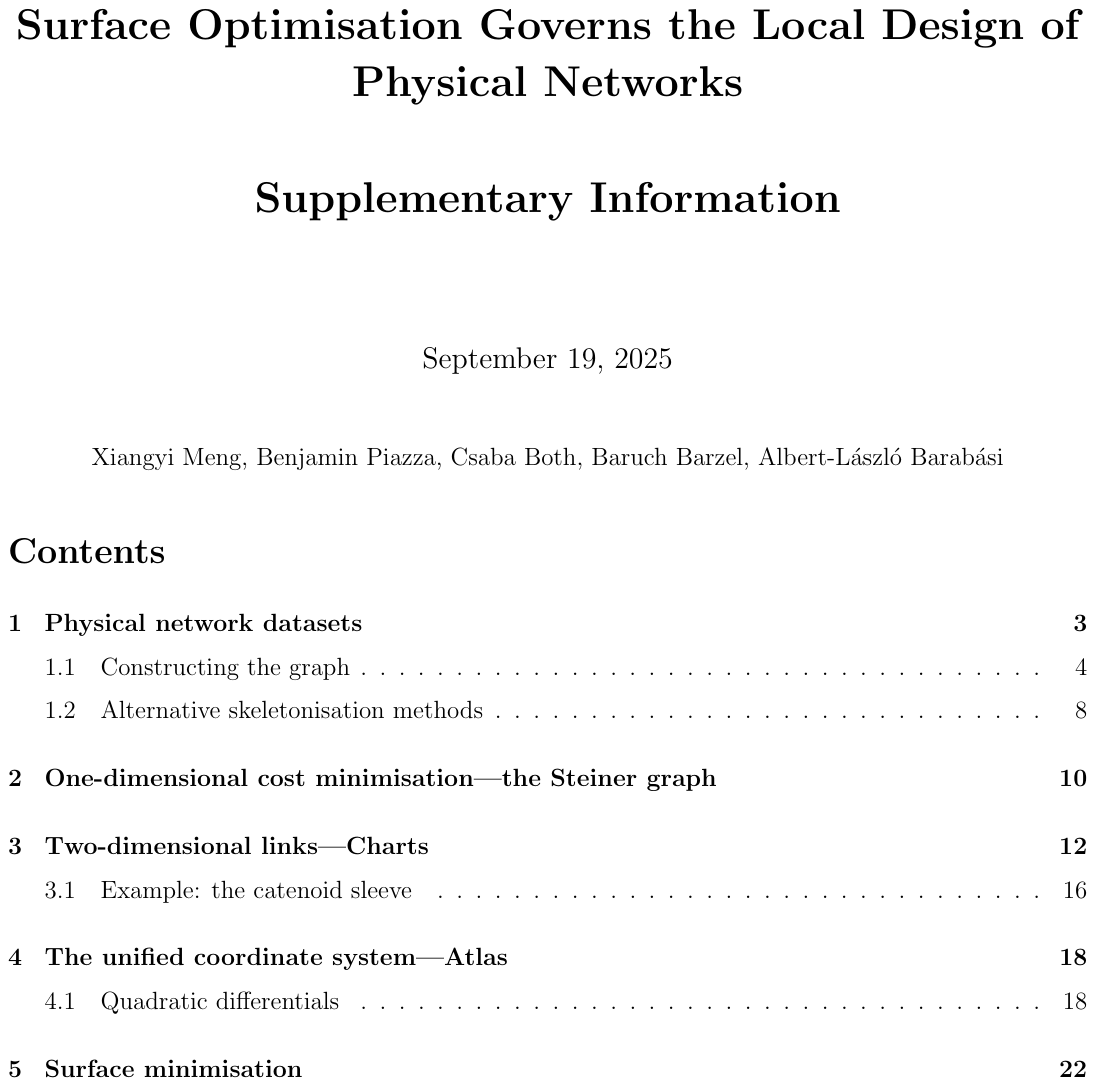}

\end{document}